\documentclass{ifacconf}

\usepackage{graphicx}      
\usepackage{natbib}        
\usepackage{amsmath}
\usepackage{amssymb}
\usepackage{subfigure}
\usepackage{multirow}
\usepackage{xcolor}
\begin{document}
\begin{frontmatter}

\title{Model-free control for machine tools} 

\thanks{This work is partially supported by the joint CNRS (France)-CSIC (Spain) PICS Program through the project UbiMFC (CoopIntEer 203 073)}

\author[First]{Jorge Villagra} 
\author[Second,Fourth]{C\'{e}dric Join} 
\author[First]{Rodolfo Haber}
\author[Third,Fourth]{Michel Fliess}

\address[First]{Centre for Automation and Robotics (CSIC-UPM), 28500 Arganda del Rey, Spain (e-mail: \{jorge.villagra,rodolfo.haber\}@csic.es)}
\address[Second]{CRAN (CNRS, UMR 7039), Universit\'e de Lorraine, BP 239, 54506 Vand{\oe}uvre-l\`{e}s-Nancy, France,  (e-mail: cedric.join@univ-lorraine.fr)}
\address[Third]{LIX (CNRS, UMR 7161), \'Ecole polytechnique, 91128
Palaiseau, France, (e-mail: Michel.Fliess@polytechnique.edu)}
\address[Fourth]{AL.I.E.N., 7 rue Maurice Barr\`{e}s, 54330 V\'{e}zelise, France,\\
       (e-mail: \{cedric.join, michel.fliess\}@alien-sas.com)}

\begin{abstract}                
Cascade P-PI control systems are the most widespread commercial solutions for machine tool positioning systems. However, friction, backlash and wearing effects significantly degrade their closed-loop behaviour.  This works proposes a novel easy-to-tune control approach that achieves high accuracy trajectory tracking in a wide operation domain, thus being able to mitigate wear and aging effects.
\end{abstract}


\end{frontmatter}

\section{Introduction}

Current automated machine tools requires high-accuracy positioning of their working axes. Several mechanical effects, often hard to identify, may compromise the appropriate positioning of the machine end-tool, thus degrading the finishing quality. To ensure that tolerances are maintained, the machine drives are equipped with tracking controllers that aims at efficiently compensate the nonlinear behaviour of the axes. 

State-of-the-art axis-positioning solutions use P and PI cascade controllers with additional feedforward compensation (\cite{armstrong1994survey}), used to counteract nonlinear effects, such as friction or backlash. In most of these compensation schemes [ranging from observers (\cite{huang2009precision}), to nonlinear identification (\cite{merzouki2007backlash}) or to evolutive algorithms (\cite{guerra2019digital}), the friction model parameters are considered constant and characterized with offline identification experiments. These models significantly degrade in the presence of additional wear-related effects. As a result, the linear control loops need to be frequently re-tuned, leaving margin for more efficient strategies.

The need for achieving nominal performance even in the presence of increased friction, motivates the investigation of nonlinear control strategies. Gain-scheduling (\cite{van2008performance}), sliding-mode (\cite{jin2009practical}), backstepping (\cite{zhang2014adaptive})  and nonlinear adaptive controllers (\cite{papageorgiou2018friction}) can eventually minimize this performance deterioration, but at the expense of a significant design complexity. In addition to that, most of these techniques focus on stability of the closed-loop dynamics without emphasizing high-accuracy positioning. 

The contribution of this paper is to present a novel control approach that attempts to answer the aforementioned challenges: (i) achieve high accuracy trajectory tracking, while (ii) keeping an easy-to-tune design, and (iii) being able to mitigate the wear and aging effects in the closed-loop behaviour. To that end, model-free control techniques, introduced in (\cite{fliess2013model}) and successfully deployed in a wide diversity of concrete case-studies\footnote{Some applications are patented.} (see e.g. \cite{fliess2013model} and \cite{bara2018} and the references therein), 
will be implemented and tested.

The outline of the paper is as follows. Section \ref{sec:system} describes the system model to be controlled, with particular emphasis on the wear-related parameters and effects. The novel control strategy is presented in Section \ref{sec:model_free}, after which some selected experimental results are showed in Section \ref{sec:results}. Finally, some concluding remark and hints on the future work are drawn in Section \ref{sec:conclusion}.


\section{System description}\label{sec:system}

\begin{figure}[thpb]
	\centering
	\includegraphics[width=0.92\linewidth]{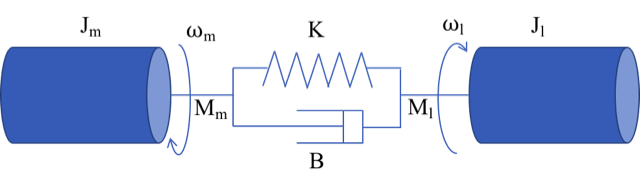}
	\caption{Scheme of machine tool axis dynamics: double mass oscillator }
	\label{figurelabel:2masses}
\end{figure}

The behavior of a machine tool axis can be represented by a double mass oscillator (motor and load) interconnected by a spring and a damper, as shown in Fig. \ref{figurelabel:2masses}. 

The dynamics of this model are parametrized by the drive motor and generalized inertias ($J_m$, $J_l$ respectively), the spring constant corresponding to the shaft stiffness $K$ and  the damping coefficient of the shaft $B$. The electro-mechanical torque and angular speed generated by the drive motor are denoted  $M_m$ and $\omega_m$, respectively, whereas the inter-connecting torque and the angular speed of the load are respectively $M_l$ and $\omega_l$.

The following transfer function describes the relationship between the motor rotation speed and the applied motor torque in the operational domain:
\begin{eqnarray}\label{eq:omegam}
\omega_m(s)&=&B_s(s)M_m(s) =\nonumber\\
&=&\dfrac{1}{J_m s}\cdot\dfrac{s^2+2D_1\omega_{0_{1}}s+\omega_{0_{1}}^2}{s^2+2D_2\omega_{0_{2}}s+\omega_{0_{2}}^2}M_m(s) 
\end{eqnarray}

\noindent where $\omega_{0_{1}}=\sqrt{\dfrac{K}{J_l}} $, $\omega_{0_{2}}=\omega_{0_{1}}\sqrt{1+\dfrac{J_l}{J_m}} $, $D_1=\dfrac{B\omega_{0_{2}}}{2K} $ and $D_2 = \dfrac{B}{2\omega_{0_{1}}J_l}$. The dynamics of this system depend essentially on the inertia of the motor and the load, as well as on the configuration of the spring-damper system. Note that in the absence of friction the damping coefficients ($D_1$, $D_2 $) and the natural frequencies ($\omega_{0_{1}}$, $\omega_{0_{2}}$) are interrelated:
{\small
\begin{equation*}
\omega_{0_{1}}=2\pi f_1,   \omega_{0_{2}}=\left(\dfrac{J_m+J_l}{J_m}\right)^{1/2}\omega_{0_{1}}, D_2=\left(\dfrac{J_m+J_l}{J_m}\right)^{1/2}D_1
\end{equation*}}

and that we can consider $f_1=\dfrac{\omega_{0_{1}}}{2\pi}$ and $D_1$ as the independent parameters which influence the whole dynamics of the system. The aging and wear effects will be modelled so that these 2 variables can take values in a broad operational domain, representative of commercially available machines nowadays: $30 \leq f_1\leq 70$, $0.08 \leq D_1 \leq 0.15$  

The relationship between the rotation speed at the load $\omega_l$ and at the motor $\omega_m$ can also be expressed in the operational domain as follows:
\begin{equation}\label{eq:omegal}
\omega_l(s)=C_s(s)\omega_m(s)=\dfrac{2D_1\omega_{0_{1}}s+\omega_{0_{1}}^2}{s^2+2D_1\omega_{0_{2}}s+\omega_{0_{1}}^2}\omega_m(s)
\end{equation}

Furthermore, the current $i_r$ can be connected with the motor torque $M_m$ while neglecting the dynamics of the electrical system with the following expression:
\begin{equation}\label{eq:Mm}
M_m(s)=A_s(s)i_r(s)=\dfrac{K_t}{J_m s}i_r(s)
\end{equation}

\noindent where $K_t$ is the electric torque constant. The current at the motor $i(t)$ has losses compared to the one generating torque $i_r$, which can be written as follows:
\begin{equation}\label{eq:ir}
i_r(t)=i(t)-i_f(t)
\end{equation}

\noindent where  $i_f$ expresses the current needed to overcome the friction  as a function of the load angular speed:
\begin{equation}\label{eq:if}
i_f(t)=D_s(\omega_l(t))= \dfrac{1}{K_t}\left( F_c sgn(\omega_l(t))+F_v\omega_l(t)\right)
\end{equation}

\noindent being $F_c$ and $F_v$ the Coulomb and viscous friction coefficients, respectively. Note also the existence of a backlash effect on the load which can play a significant role during the reversal phases of the control signal. 

\begin{figure*}[thpb]
	\centering
	\includegraphics[width=0.91\linewidth]{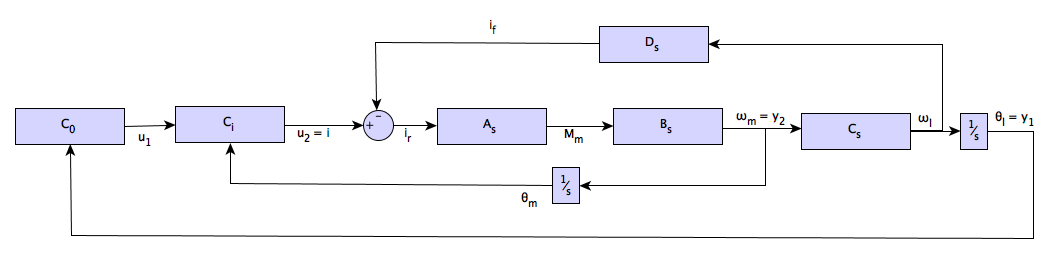}
	\caption{Block diagram of the machine tool control structure }
	\label{figurelabel:schemabloc}
\end{figure*}

Fig. \ref{figurelabel:schemabloc} depicts the dependencies between expressions (\ref{eq:omegam}) - (\ref{eq:if}) and their interactions with the two control loops $C_0,C_1$which are detailed in the following section.

\section{Model-free approach for drive-trains}\label{sec:model_free}

\subsection{Cascade P-PI control}
A widely accepted control structure in the industry is a cascade with 2 loops: (i) an external one on the position of the load, where typically a proportional corrector is used, and (ii) an internal one on the speed of the motor, where a PI is often implemented. The outer loop $u_1$ is closed using a feedforward term $C_{0_{ff}}$ and a  feedback term $C_{0_{fb}}$:
\begin{eqnarray}\label{eq:u1_PPI}
u_1(t)&=&C_0\left(\theta_l(t)\right)=C_{0_{ff}}(t)+C_{0_{fb}}(\theta_l(t))=\nonumber\\ 
&=& \dot{\theta}_l^*(t)+\Phi_1\left(\theta_l^*(t)-\theta_l(t)\right)
\end{eqnarray}

\noindent where $\theta_l^*(t)$ is the load position reference trajectory and $\Phi_1$ is a generic corrector applied to the load position error. The inner control loop  is also a combination of feedforward  $ C_ {1_ {ff}} $ and feedback $C_{1_{fb}}$:
{\small
\begin{eqnarray}\label{eq:u2_PPI}
u_2(t)&=&C_1\left(\omega_m(t)\right)=C_{1_{ff}}(t)+C_{1_{fb}}(\omega_m(t)))=\nonumber\\
&=&\dfrac{Jm+Jl}{Kt}\dot{\omega}_l^*(t)+\Phi_2\left(\omega_m^*(t)-\omega_m(t)\right),\; \omega_m^*(t) = u_1(t)
\end{eqnarray}}

\noindent where $\omega_m^*(t) $ is the reference trajectory for the motor speed which, given the cascade structure of Fig. \ref{figurelabel:schemabloc}, is equal to the external loop control variable $u_1(t)$. The term $\Phi_2 $ represents a generic corrector applied to the engine speed error. Note that in the case of a P-PI scheme $\Phi_1=K_{p_{o}}e$ and $\Phi_2=K_{p_{i}}e+K_{i_{i}}\int e dt$

\subsection{Model-free control principles}

Model-free controllers are used in this work because they combine the well-known and easy-to-tune PID structure with an ``intelligent'' term that compensates the effects of nonlinear dynamics, disturbances  or uncertain parameters. 

As demonstrated in (\cite{fliess2013model}), most SISO systems can
be written locally as
\begin{equation} \label{eq:sist_mod}
\dot{y} = F + \alpha u
\end{equation}
\noindent where $\alpha \in \mathbb{R}$ is a constant parameters, which do not necessarily represent a physical magnitude, and whose value is chosen by the practitioner such that
it allows $F$ and $\alpha u$ to be of the same order of magnitude.

The data-driven term $F$, which includes not only the unknown structure of the system but also any disturbance (\cite{fliess2013model}), is computed as follows:
\begin{equation}
{\displaystyle \hat{F}(t_k)=\hat{\dot{y}}(t_k)-\alpha u(t_{k-1})}
\end{equation}

Taking the above into consideration, the loop can be closed with an intelligent controller (iP) using the following expression:
\begin{equation}\label{eq:iP_gen}
u=K_pe+\dfrac{\dot{y}^*-\hat{F}}{\alpha}
\end{equation}

\noindent where $y^*$ is the reference trajectory, $e=y-y^*$ is the tracking error and $K_P\in\mathbb{R}$ is a gain. Note that the tuning complexity of this approach is comparable to a PI controller, as only 2 parameter need to be chosen.

\subsection{Cascade model-free control}

The classic P-PI structure is replaced by another scheme based on a iP-iP structure, where the following outer and an inner input-output model are used (see \cite{lafont2015model} for an explanation for such MIMO systems):
\begin{equation*}
\dot{y}_1=F_1+\alpha_1u_1,\;
\dot{y}_2=F_2+\alpha_2u_2
\end{equation*}

\noindent where inputs $u_1, u_2 $ and outputs $y_1=\theta_l, y_2=\omega_m$ corresponds to the signals represented  in Fig. \ref{figurelabel:schemabloc}, and $\alpha_1,\alpha_2\in\mathbb{R}$ are gains chosen by the control engineer.

Following the expression of a generic iP presented in (\ref{eq:iP_gen}), the outer and inner loops are respectively closed with feedback controllers $\Phi_1^*$ and $\Phi_2^*$ to which feedforward terms $C_ {0_ {ff}}=\dot{\theta}_l^*$ and $C_ {1_ {ff}}=\dfrac{Jm+Jl}{Kt}\dot{\omega}_l^*$, introduced in (\ref{eq:u1_PPI}) and (\ref{eq:u2_PPI}), are respectively added as follows: 
\begin{eqnarray}\label{eq:u1_iPiP}
&u_1(t)=C_0\left(\theta_l(t)\right)= \dot{\theta}_l^*(t)+\Phi_1\left(\theta_l^*(t)-\theta_l(t)\right),\nonumber\\ &\Phi_1^*=K_{p_{o}}^*\left(\theta_l^*(t)-\theta_l(t)\right)+\dfrac{1}{\alpha_1}\left(\dot{\theta_l^*}-\hat{F}\right)
\end{eqnarray}

\begin{eqnarray}\label{eq:u2_iPiP}
&u_2(t)=C_1\left(\omega_m(t)\right)=\dfrac{Jm+Jl}{Kt}\dot{\omega}_l^*(t)+\Phi_2\left(\omega_m^*(t)-\omega_m(t)\right)\nonumber\\
&\Phi_2^*= K_{p_{i}}^*e_m+\dfrac{1}{\alpha_2}\left(\dot{\omega_m^*}-\hat{F}\right),\nonumber\\
&\e_m=\omega_m^*(t)-\omega_m(t),\; \omega_m^*(t) = u_1(t)
\end{eqnarray}

\noindent  where $K_{p_{o}}^*$ and $K_{p_{i}}^*$ are  the proportional gains of the outer and inner iP, respectively. 


\section{Experimental results}\label{sec:results}

\begin{figure}[tbp]
    \centering
        \includegraphics[width=1.05\linewidth]{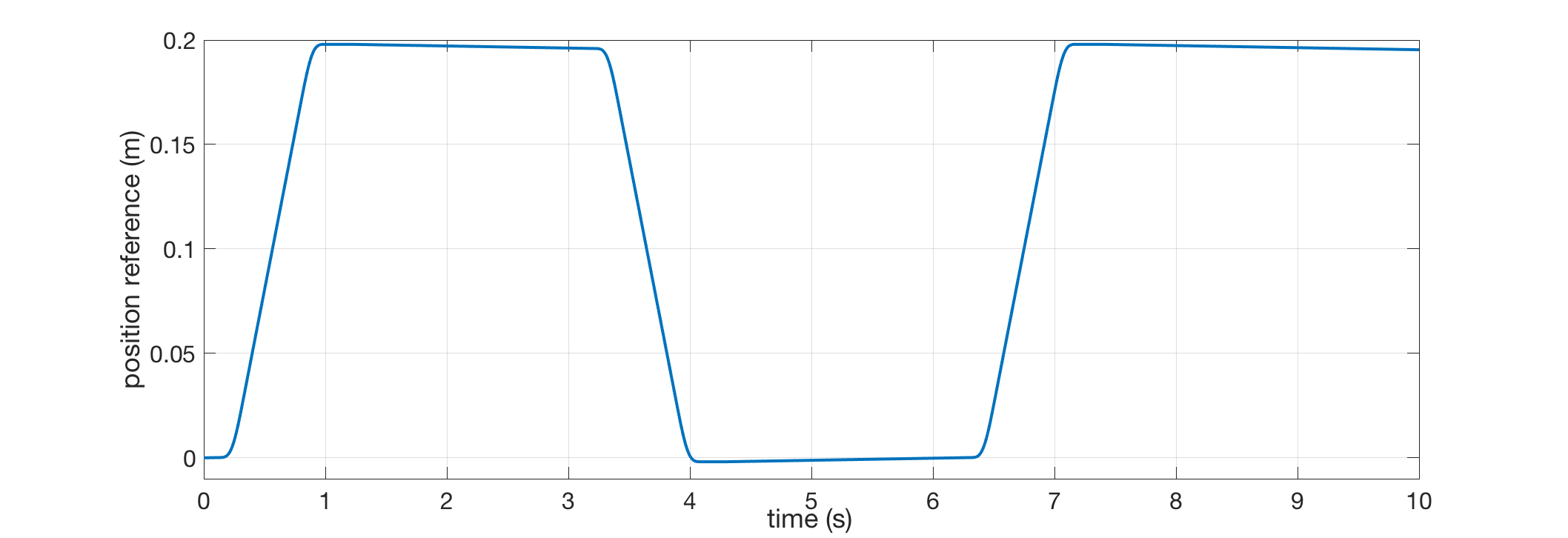}
       \includegraphics[width=1.05\linewidth]{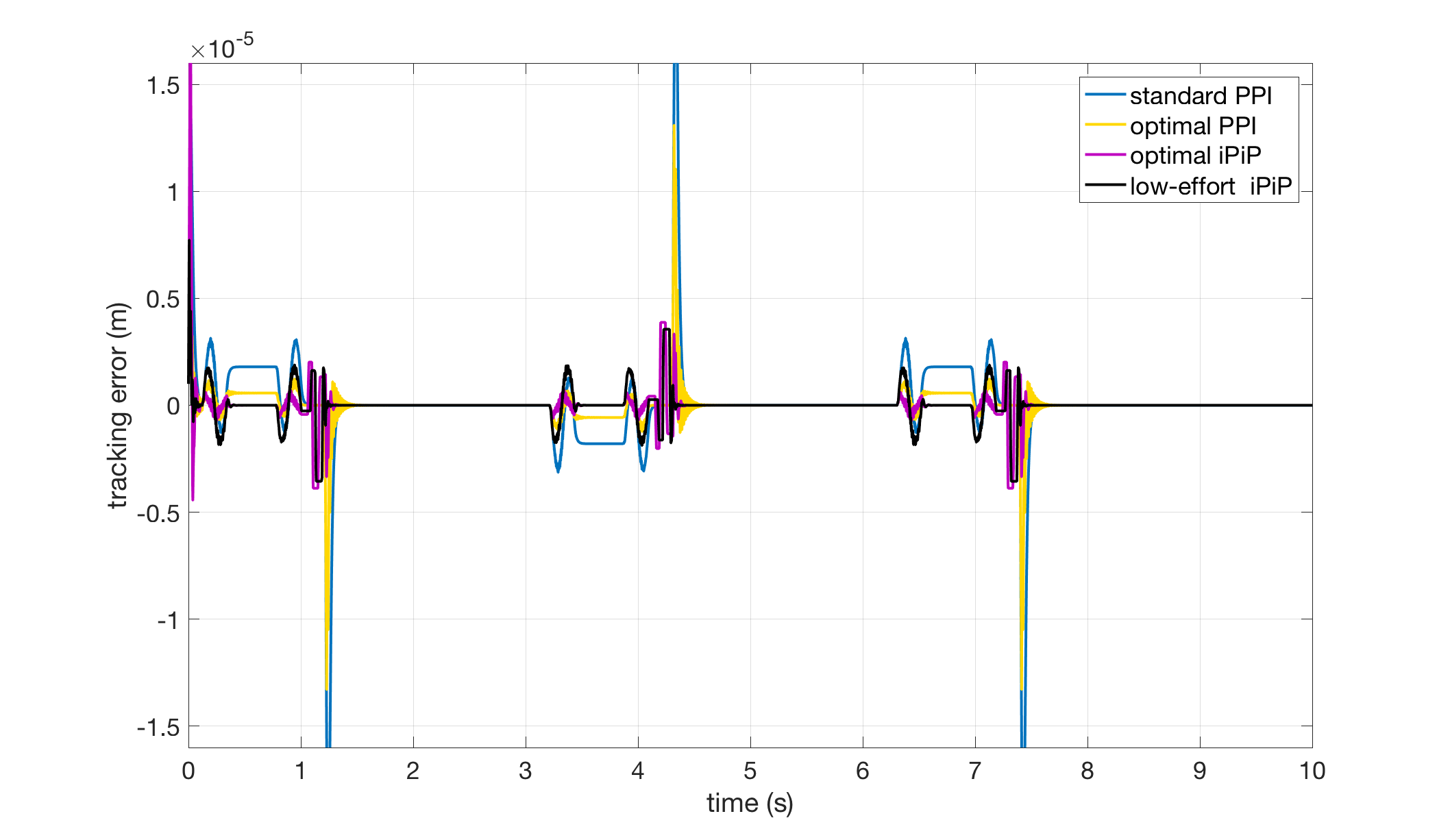}
	\caption{a) Load reference trajectory;  b) Tracking error}  
    \label{fig:results_comp}   
\end{figure}

The P-PI cascade controller and the model-free iP-iP control system, expressed respectively in (\ref{eq:u1_PPI})-(\ref{eq:u2_PPI}) and in (\ref{eq:u1_iPiP})-(\ref{eq:u2_iPiP}) have been thoroughly compared. To that end, a benchmark reference trajectory has been selected (see Fig \ref{fig:results_comp}a), where several inversions zones challenge the control system.

\begin{table*}
\centering
\caption{Comparison of PPI and iPiP under different operation conditions}
\begin{tabular}{|l|c|c|c|c|c|c|}
\cline{3-7}
\multicolumn{2}{l|}{} &  P-PI nom     & P-PI nom opt & iP-iP        & iP-iP+FF      & iP-iP+FF (wrong param.) \\\hline
\multirow{2}{*}{$\Sigma_1(D_1=70,f_1=0.15)$} & ITAE & $2.978\cdot10^{-5}$ & $1.024\cdot10^{-5}$ &  $\color{blue}\mathbf{6.823\cdot10^{-6}}$ & $9.554\cdot10^{-6}$ & $9.598\cdot10^{-6}$ \\ \cline{2-7}
                  & IAU & 5.360 & 5.366 & 6.221 & $\color{blue}\mathbf{5.162}$ & 5.162 \\\hline
\multirow{2}{*}{$\Sigma_2(D_1=30, f_1=0.08)$} & ITAE & $7.319\cdot10^{-2}$ & $3.134\cdot10^{-4}$ & $\color{blue}\mathbf{1.680\cdot10^{-4}}$ & $1.913\cdot10^{-4}$ & $1.944\cdot10^{-4}$ \\\cline{2-7}
                  & IAU  & 119.6 & 5.731 & 5.456 & $\color{blue}\mathbf{5.343}$ & 5.345 \\\hline
\end{tabular}\label{table:PPI_vs_iPiP}
\end{table*}

To quantitatively compare both control approaches, the following 2 key performance indicators have been chosen:
\begin{equation*}
   ITAE = \int_0^T t|\theta_l-\theta_l^*|dt, \;
   IAU = \int_0^T |u|dt
\end{equation*}

\noindent which will be computed considering the whole testing interval $t\in[0,10]$. 

Fig. \ref{fig:results_comp}b and Table \ref{table:PPI_vs_iPiP} allow to see and quantify the behaviour of each strategy. In both elements, 4 different control configurations have been analysed for a system with significantly different dynamic behaviour (in configuration $\Sigma_1$, $D_1=70,f_1=0.15$, while in  $\Sigma_2$, $D_1=30,f_1=0.08$):

\begin{enumerate}
\item a P-PI, whose gains ($K_{p_{o}}, K_{p_{i}}, K_{i_{i}} $) are obtained from a commercial control system for machine tools 
\item a P-PI, whose gains ($K_{p_{o}}, K_{p_{i}}, K_{i_{i}}$) have been optimised with respect to criteria $\mathcal{J}=ITAE +w_u*IAU$ for a specific operation condition 
\item an iP-iP, whose gains ($\alpha_1, \alpha_2, K_{p_{o}}^*, K_{p_{i}}^*$) have been optimized with respect to the same criteria and for the same specific operation condition 
\item an iP-iP identically tuned, but incorporating a model-based feedforward (FF).
\end{enumerate}

Note that the last control variant introduces an anticipatory control term:
\begin{equation}
    C_{ff}(s)=\dfrac{s^2+2D_2\omega_{0_{2}}s+\omega_{0_{2}}^2}{s^2+2D_1\omega_{0_{1}}s+\omega_{_{1}}^2}\omega_l^*(s)
\end{equation}

\noindent which presumes $D_1$ and $f_1$ well know, which is not often the case, unless off-line identification tests have been conducted. The motivation to include this term is to reduce the control effort generated by iP-iP controllers. However, as the involved parameters may be badly known, a fifth item to be compared has been introduced in Table \ref{table:PPI_vs_iPiP}, aiming at assessing the sensitivity of the closed-loop behaviour to wrong values of $D_1$ and $f_1$ ($0.15$ and $70$ instead of $0.08$ and $30$, and viceversa).

As can be observed, the iP-IP controller achieves a significant improvement both with respect to the standard and the optimised P-PI, both quantitatively -see ITAE- and qualitatively -lower inversion peaks. Although the control action is higher in the regular iP-iP control, the inclusion of an anticipative model not only mitigates this aspect, but it even achieves a lower tracking error than P-PI.

\begin{figure}[thpb]
	\centering
	\includegraphics[width=0.75\linewidth]{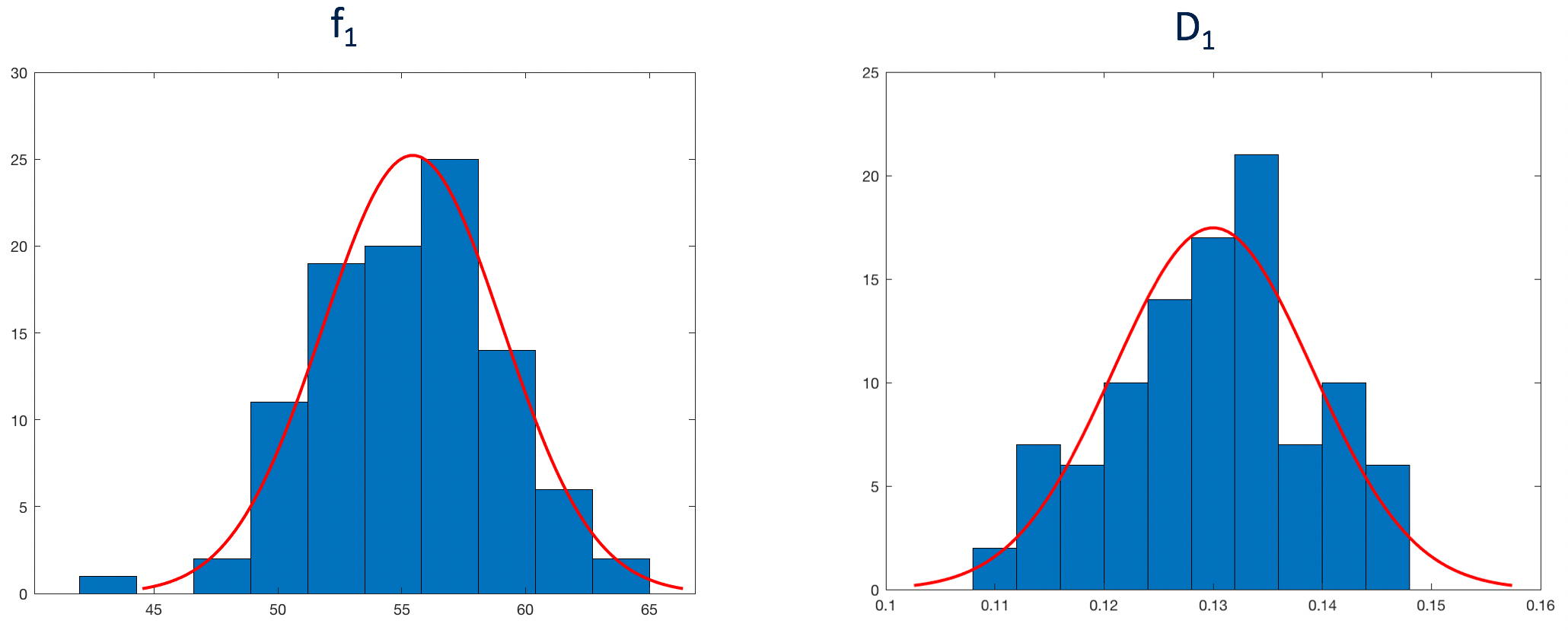}
	\caption{Histograms with generated values of $f_1$ and $D_1$}
	\label{fig:param_distrib}
\end{figure}

A key consideration of this work is the assessment of model-free controllers under a significant variation of wear related-parameters, namely $D_1$ and $f_1$. To that end, a Monte-Carlo simulation has been conducted using normal distributions of such parameters, as depicted in Fig. \ref{fig:param_distrib}, where the desired operating domains have been approximated by conservative normal distributions $f_1\sim \mathcal{N}(55,4)$ and $D_1\sim\mathcal{N}(0.13,0.01)$.

 The existing control approaches have difficulty in obtaining good behavior in these ranges, which limits the field of operation to a very restricted set of machines. 

\begin{figure}[thpb]
	\centering
	\includegraphics[width=\linewidth]{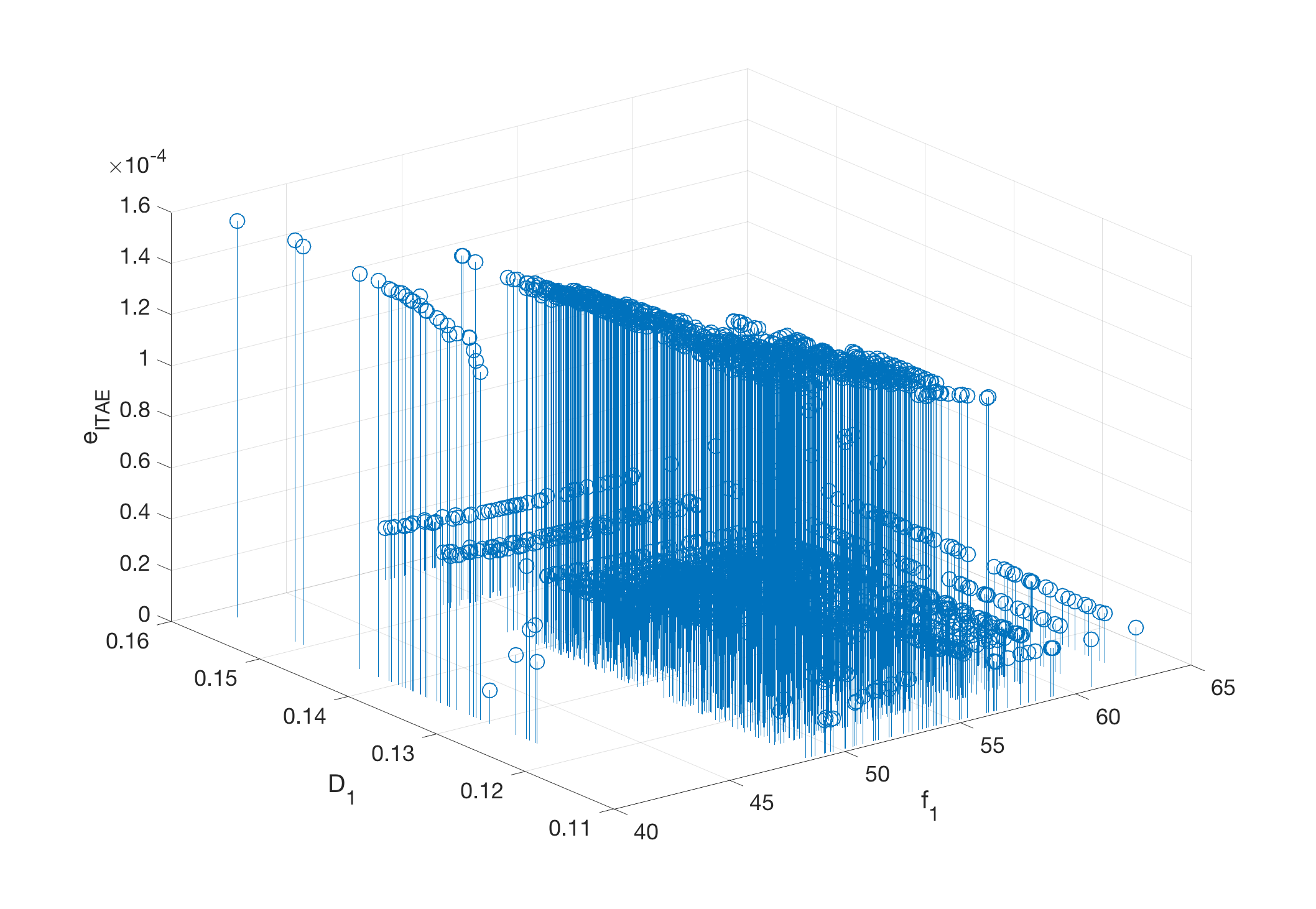}
	\caption{Stem representation of the Monte-Carlo test, with $f_1\sim \mathcal{N}(55,4)$ and $D_1\sim\mathcal{N}(0.13,0.01)$}
	\label{fig:stem_montecarlo}
\end{figure}

Fig. \ref{fig:stem_montecarlo} represents the difference in terms of ITAE between the standard P-PI and the iP-iP control structures for the generated input parameter combinations. The values are positive in every tested case, showing that iP-iP provides also more accurate positioning than P-PI when wearing effects appear.

\section{Concluding remarks}\label{sec:conclusion}

An easy-to-tune model-free control approach has been presented for axis-positioning in machine tool systems.
The preliminary results of this work exhibits an outstanding tracking behaviour not only under a specific operation condition, but also when a significant wear-induced parameter range is considered. 



%
%
%
%
%
                                      
\end{document}